\documentstyle[aps,epsf]{revtex}

\begin{document}

\draft

\twocolumn[\hsize\textwidth\columnwidth\hsize\csname@twocolumnfalse\endcsname

\title{$c$ axis superfluid response of Copper-Oxide
superconductors}

\author{T. Xiang and J. M. Wheatley}

\address{Research Centre in Superconductivity, University of
Cambridge, Madingley Road, Cambridge CB3 OHE, United
Kingdom.}

\date{\today}

\maketitle

\begin{abstract}

A novel interplay between $d$-wave superconducting order
parameter symmetry and the underlying Cu $3d$ orbital based
electronic structure of Copper-Oxides leads to a striking
anisotropy in the superfluid response of these systems.  In
clean tetragonal materials the $c$-axis penetration depth
increases as $T^5$ at low temperature $T$, in contrast to
linear $T$ behavior in the $ab$-plane.  Disorder is a
relevant perturbation which causes all components of the
superfluid response to depend quadratically on temperature
at low temperature.  However, the cross-over temperature
scale from the intrinsic $d$-wave behavior to the disorder
dominated behavior for the in-plane response may be
different from that for the out-plane response.

\end{abstract}


]


A variety of experimental data on Copper-Oxide
Superconductors are compatible with a pairing state of
$d_{x^2-y^2}$ symmetry in CuO$_2$ planes.  The existence of
energy gap nodes in this state has profound consequences for
the low temperature electro-magnetic response.  The in plane
penetration depth increases linearly with $T$ at low $T$ in
contrast to the activated behavior of conventional
superconductors.  The linear-$T$ result holds in the regime
where the response is governed by coherently propagating
fermion quasi-particle excitations, i.e.  at temperatures $
\Gamma \ll T \ll T_c$ where $\Gamma$ is the superconducting
quasiparticle scattering rate.  Linear-$T$ behavior in the
$ab$-plane has now been established in both chain and
non-chain Copper-Oxides\cite{Hardy93,Jacobs95,Lee96,Panagopoulos96}.
Experiment\cite{Hardy93,Jacobs95,Panagopoulos96,Shibauchi94,Morgan},
however, indicates a much weaker T dependence of c-axis
penetration depth.  Straightforward extension of the
$d$-wave model to 3D leads to line gap nodes on a warped
cylindrical Fermi surface.  This gives a linear-$T$ $c$-axis
response in apparent contradiction to
experiment\cite{Radtke96}.

In this paper we propose that weak $T$-dependence of the
$c$-axis superfluid density $\rho_s^c$ is a general feature
of non-chain cuprates with crystal tetragonal symmetry,
whose origin lies in the simple but unusual electronic
structure of these materials.  Our analysis is based on the
assumption that coherent Bloch bands along $c$-axis are
present at low temperatures, i.e.  $k_z$ is a good quantum
number.  As discussed below, an essential feature of the
electronic structure is that the $c$-axis hopping integral
$t_\perp (k_\parallel)$ is a function of the in plane
momentum $k_\parallel$ and is vanishingly small when
$k_\parallel$ lies along the zone diagonals of the 2D
Brillouin zone\cite{novikov93,Andersen95}.  The nodal lines of the
$d_{x^2-y^2}$-wave gap $\Delta (k_\parallel)$ and the zeros
of $t_\perp(k_\parallel)$ therefore coincide.  This weakens
the $d$-wave node contribution along the $c$-axis relative
to the $ab$-plane and leads to a weak $T$-dependence of the
$c$-axis superfluid response $\rho_{s}^c$.  At sufficiently
low temperatures disorder effects are always relevant, and
tend to enhance the $T$ dependence of $\rho_s^c$.

Let us first consider the electron structure in high-$T_c$
compounds with tetragonal symmetry.  For simplicity we
consider a tetragonal monolayer system, for example ${\rm Hg Ba_2 Cu
O_{4+\delta} }$\cite{crystals}.  LDA band structure
calculations reveal that a minimal tight-binding model for
high-$T_c$ compounds involves nearest-neighbour hopping
between Cu $3d_{x^2-y^2}$ and $4s$ and O $p_x$ and $p_y$
orbitals\cite{novikov93,Andersen95}.  Thus a 
model Hamiltonian for ${\rm Hg Ba_2 Cu O_{4+\delta} }$,
expressed in terms of bonding, $\alpha_k =2 (\cos k_x {\rm
p}_x + \cos k_y {\rm p}_y) / \omega_k$ where $\omega_k = 2
\sqrt{\cos k_x^2 + \cos k_y^2}$, and non-bonding, $\beta_k
=2 ( \cos k_y {\rm p}_x - \cos k_x {\rm p}_y ) / \omega_k$,
representation of orthogonal oxygen Wannier orbitals with
respect to the Cu $d_{x^2-y^2}$ orbital\cite{srep}, is given 
by
\begin{eqnarray} 
H &=& \sum_k C^\dagger_k \left(\begin{array}{cccc} 
-E_d & 0 & t_{pd}\omega_k & 0 \\ 0 &
E_s+ 2 t_{ss} d \cos k_z & t_{ps}\mu_k & t_{ps}z_k \\
t_{pd}\omega_k & t_{ps}\mu_k & 0 & 0 \\ 0 & t_{ps}z_k & 0 &
0 \end{array}\right) C_k \nonumber\\
&& + \sum_i U n_{di}^2 ,\label{model}
\end{eqnarray} 
where $C_k = (d_k,\, s_k,\, \alpha_k ,\, \beta_k )$,
$(E_d,\, E_s)$ are the energy of $(3d,\, 4s)$ orbital with
respect to the O orbital, and $(t_{pd}, \, t_{ps}, \,
t_{ss})$ are wavefunction overlap integrals among $d$, $s$,
and $p$ orbitals.  The symmetry functions appearing here are
$\mu_k = 2 (\cos k_x - \cos k_y)/\omega_k$ and $z_k = 8 \cos
k_x \cos k_y/\omega_k$.  The Hamiltonian (\ref{model})
differs from that in Ref.  \cite{Andersen95} purely by
retention of a strong Coulomb interaction $U$ in the Cu $d$
orbital.   This produces a charge transfer insulator at
half filling, a feature absent in LDA calculations.

In the strong coupling limit, the four band model
(\ref{model}) can be simplified as a one-band model by
eliminating the highlying Cu $4s$ orbitals and the highlying
$d-p$ spin triplets.  This projection procedure can be
carried out in two steps:  firstly, by elimination of the
highlying $4s$ orbital; and secondly, by solving the
correlation problem within the unit cell and treating
intercell hopping as a degeneracy lifting perturbation
following Refs.  \cite{Shastry89,Mattis95}.  This yields the
well-known t-J model with both intra- and interlayer
hoppings and exchanges.  We assume the non-bonding orbitals
are sufficiently lowlying that they are filled and inert.  We
notice that the dominant $c$-axis overlap is via large
radius $s$ orbitals \cite{Andersen95}.  Thus terms to second
order in $E_s^{-1}$ must be retained to describe effective
$c$-axis hopping via O bonding orbitals.  The lowest energy
states within the $CuO_2$ unit cell are spin 1/2 doublets $|
1 \sigma \rangle$ with one hole per unit cell and Zhang-Rice
singlets $| 2 \rangle$ with two holes per unit cell.  The
$ab$-plane hopping matrix elements for Zhang-Rice singlets
involve both $d$-$\alpha$ and effective $\alpha$-$\alpha$
overlap.  However along the $c$-axis, the only channel for
hopping is via the bonding O orbitals.  It can be shown that
the projected $c$-axis hopping integral has the form
$t_\perp (k_\parallel ) =t_\perp^0 \mu_k^2$, where
$t_\perp^0 = t_{ss} (t_{pd}t_{ps}M_0 / E_dE_s)^2$ and $M_0 =
\langle 1 \sigma | \alpha_\sigma^{\dag} |2 \rangle$ is the
overlap of the state produced by eliminating the hole in the
O bonding orbital and the ground state doublet.  Thus
$t_\perp (k_\parallel )$ contains a $k_\parallel$ dependent
factor $\mu_k^2$, which vanishes when $k_x = \pm
k_y$\cite{ab-plane}.  This is a robust feature of realistic
strong correlation models of non-chain tetragonal materials
and directly reflects the matrix element between Cu 4$s$ and
bonding O $\alpha$ orbitals.  It is worth pointing out that
$ab$-plane anisotropy observed in $c$-axis magneto-transport
experiments may be dominated by precisely this
effect\cite{hussey96}.

\begin{figure}
\leavevmode\epsfxsize=12cm
\epsfbox{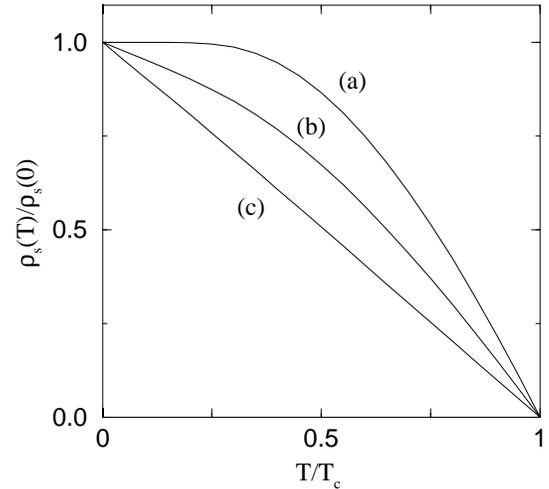}
\caption{Normalized superfluid density $\rho_s(T)/ \rho_s 
(0)$ vs $T/T_c$. The energy dispersion of electrons  $
\varepsilon_k = - 2 t (\cos k_x + \cos k_y) - 4 t^\prime 
\cos k_x \cos k_y - 2 t_\perp (k_\parallel ) \cos k_z 
$ is used in the calculation. 
 (a) is for the model with $t_\perp (k_\parallel )
= t_\perp^0 (\cos k_x - \cos k_y )^2 / 4$  along $c$-axis. 
(b) and (c) are for the model with $t_\perp (k_\parallel ) = t_\perp^0$ 
along $c$- and $a$-axis, respectively.  
$t=1$, $t^\prime =-0.25$, $t_\perp^0 =0.1$, $T_c =0.14$, 
and the filling factor is 0.4.  (The results are not sensitive 
to the change of parameters). }  
\label{fig1}
\end{figure}

Now let us consider the effect of this $k_\parallel$
dependent $t_\perp$ on the $c$-axis superfluid response.
For a clean system the electromagnetic response is governed by
coherent 3D fermion quasiparticle excitations at low
temperature.  In this case the superfluid density (or
inverse square of penetration depth) is given
by\cite{Scalapino92}
\begin{equation}
	\rho^\mu_s = \sum_k \left[ 2 \left({\partial 
	\varepsilon_k\over \partial k_\mu }\right)^2 {\partial 
	f(\lambda_k)\over \partial \lambda_k} -{\partial^2 
	\varepsilon_k\over \partial k_\mu^2}{\varepsilon_k 
	\over \lambda_k } \tanh (\beta\lambda_k /2)\right] ,
	\label{superfluid}
\end{equation}
where $\varepsilon_k$ is the energy dispersion of electrons,
$\lambda_k = \sqrt{ \varepsilon_k^2 + \Delta_k^2}$, and
$f(\lambda_k)$ is the Fermi function.  When $t_\perp (k_\parallel 
) =t_\perp^0 \mu_k^2$ it is
straightforward to show that at low temperature the $c$-axis
superfluid response is
\begin{eqnarray}
	\rho^c_s(T) &\sim  &{3\over 4}N(0) (t_\perp^0)^2 \left( 
	1 - { 5\over 12 } \left( {\Delta (T) \over \Omega} 
	\right)^2 -450 \left( { T\over \Delta_0 } \right)^5
	\right. \nonumber\\ && 
	+ \left. o(T^5)\right) , 
	\label{T5law}
\end{eqnarray}
where $N(0)$ is the normal density of states on CuO planes,
$\Delta (T)$ is the energy gap, $\Delta_0 =\Delta (0)$, and
$\Omega$ is the energy cut off which is much larger than $\Delta$.  
The $T^5$ term is from the first term in
(\ref{superfluid}):  one $T$ is from the contribution of the
linear density of states of the d-wave state and the other
$T^4$ is due to the $(\cos k_x - \cos k_y )^4$ factor in
$t_\perp^2 (k_\parallel )$.  The second term in
(\ref{T5law}) comes from the second term in
(\ref{superfluid}), at low temperature it contributes a
positive $T^3$ term to $\rho^c_s$ since the d-wave gap
$\Delta (T) / \Delta_0 \sim 1 - (T/ \Delta_0)^3$.  However,
the coefficient of this $T^3$ term is proportional to $(
\Delta_0 / \Omega )^2$, which is extremely small.  Therefore
it is likely that the dominant temperature dependence of
$\rho^c_s$ should be $T^5$ at low temperature.  This $T^5$
behavior in $\rho^c_s$ is clearly very different from the
$t_\perp (k_\parallel ) =constant$ case where $\rho^c_s$
varies linearly with T at low temperature.  In Fig.
\ref{fig1}, we compare the temperature dependence of
$\rho^c_s (T) /\rho^c_s (0)$ in the whole temperature regime
for the above two cases.  The temperature dependence of
$\rho_s (T) /\rho_s (0)$ along $a$-axis is also shown in
Fig.  (\ref{fig1}) for comparison.

\vspace{-2cm}

There are several effects which may lead to a finite hopping
integral along the $c$-axis in the vicinity of the gap nodes
in clean systems and thus a stronger $T$ dependence of $\rho_s^c$.  
Many non-chain materials have some degree
of orthorhombicity, i.e.  $t_x \ne t_y$.  Clearly this
will affect the location of the gap nodes and the zeros of
$t_\perp(k_\parallel )$.  Following the previous treatment
for the model (\ref{model}), it can be shown that the
projected $c$-axis hopping integral is still a function of
$k_\parallel$, but $t_\perp (k_\parallel ) $ now has
approximately the form $t_\perp^0 (\cos k_x -{t_y\over t_x}
\cos k_y )^2/4$.  Thus the zeros of $t_\perp (k_\parallel )$
are no longer on $k_x =\pm k_y$ planes.  The gap nodes
may not lie on $k_x =\pm k_y$ planes either, but
there is no symmetry reason requiring the gap nodes to
coincide to the zeros of $t_\perp (k_\parallel )$.  Thus
$t_\perp$ is now finite at the gap node.  Direct interplane
O-O overlap can also lead to a finite dispersion around the
gap nodes and eliminate the zeros in $t_\perp
(k_{\parallel})$ altogether.  In either case, we may model
the $c$-axis electronic structure by $t_\perp(k_\parallel) =
t_\perp^0 \mu_k^2 + t_\perp^{node}$ with ${ t}_\perp^{node}$
the value of $t_\perp$ at the gap nodes.  Both effects are
expected to be weak, but they generate a small linear term
in $\rho_s^c$, with a slope proportional to $(t^{node}_\perp
)^2$, which dominates the $c$-axis superfluid response when
$T << t_\perp^{node}$.

Sufficiently close to a nodal line the $c$-axis hopping rate
always becomes small in comparison with the impurity
scattering rate, and thus disorder is always a relevant
perturbation.  Disorder introduces a finite quasiparticle
lifetime and consequently a finite density of states at the
Fermi level.  At low temperature the correction to the 
superfluid density caused by finite quasiparticle scattering 
rates is approximately given by 
\begin{equation}
 \delta \rho_s^\mu \sim -{2\over \pi} \sum_k \left( 
 {\partial \varepsilon_k \over \partial k_\mu } \right)^2
 \int_{-\infty}^\infty {\rm d} \omega f(\omega ) 
 {\rm Im} { (\omega + i \Gamma )^2 + \lambda_k^2 
 \over [(\omega + i \Gamma )^2 - \lambda_k^2]^2 }. 
\end{equation}
It can be shown that both the $ab$ plane and the $c$-axis
superfluid responses behave as $T^2$ when $T$ is much
smaller than the quasiparticle scattering rate
$\Gamma$\cite{lifetime}.  However, the coefficients of the
$T^2$ terms are very different for the $ab$ plane and the
$c$-axis responses.  On the $ab$ plane, $\rho^{ab}_s(T) /
\rho^{ab}_s(0) \sim - (\Delta_0 / \Gamma ) (T /
\Delta_0)^2$; while along the $c$-axis, $\rho^c_s(T) /
\rho^c_s(0) \sim - (\Gamma / \Delta_0 ) (T / \Delta_0)^2$.
Thus the $T^2$ behavior of the $c$-axis response is by a
factor $(\Gamma / \Delta_0)^2$ weaker than that on the $ab$
plane (generally $\Gamma \ll \Delta_0$) due to the $\mu_k^2$
factor in $t_\perp (k_\parallel )$.  Notice that disorder
has opposite effects in $ab$ and $c$-axis; it weakens the
$T$-dependence of the in-plane response and strengthens the
$T$-dependence of the out of plane response.

Another important aspect of disorder is that it may disrupt
the symmetry of bonding orbitals about the Copper site.  The
selection rule preventing hopping for momentum along
$k_x=\pm k_y$ is then relaxed.  For example, consider the
effect of interlayer defects on O site.  This leads to a
random component of $t_\perp$ due to random variation in the
matrix element $M_0$ for a Zhang-Rice singlet discussed
earlier.  Near the gap nodes this fluctuating component in
$t_\perp$ dominates $c$-axis hopping, and thus electrons in
the vicinity of nodes are well described by an impurity
assisted hopping model which has been introduced in the
literature on phenomenological grounds\cite{Radtke96}.
Impurity assisted hopping gives a new conduction channel and
has a direct contribution to the $c$-axis superfluid density
$\rho_{s,imp}^c$:
\begin{eqnarray}
 \rho_{s,imp}^c &\sim &{4\over \beta}\sum_{\omega_n, k, k^\prime}
 \langle \delta t_{\perp} (k ) 
 \delta t_{\perp} ( -k )\rangle_{imp} \nonumber\\
 &&{\rm Tr} \left[
 \sin^2 {k_z \over 2} G(k + k^\prime , \omega_n ) G(k^\prime , 
 \omega_n) \right. \nonumber \\
 && \left. - \cos^2 {k_z \over 2} 
 G(k+ k^\prime , \omega_n ) \tau_3 G(k^\prime ,  
 \omega_n)  \tau_3 \right]. 
\end{eqnarray}
Thus the superfluid response induced by the impurity assisted 
hopping $\rho^c_{s,imp}$  depends on the impurity average of
the assisted hopping matrix element $\langle \delta t_{\perp
}(q) \delta t_{\perp} (-q)\rangle_{imp} $.  If the impurity
scattering is isotropic in space, i.e.  $\langle \delta
t_{\perp }(q) \delta t_{\perp} (-q)\rangle_{imp} $ is
independent on $q$, then the assisted hopping contribution
to $\rho_s^c$ in a $d$-wave state is zero, resulted from the
vanishing average of the $d$-wave gap on Fermi surface.  If,
however, the scattering is anisotropic, for example if
$\langle \delta t_{\perp }(q) \delta t_{\perp}
(-q)\rangle_{imp} $ has the Lorentzian form $\langle \delta
t_{\perp }(q) \delta t_{\perp} (-q)\rangle_{imp} = {v_0^2
k_F\delta k \over ({\bf k}-{\bf k}^\prime )^2 + (\delta
k)^2}$,\cite{Radtke96} then in the strong forward scattering 
limit ($\delta
k \rightarrow 0$) $\rho^c_{s,imp}$ behaves as $T^2$ at low
temperature:
\begin{equation} 
 \rho^c_{s,imp} \propto
 \Gamma_\perp \Delta N(0) \left[ 1 -8 \ln 2 \left( {T\over
 \Delta}\right)^2 + o(T^2) \right] 
\end{equation}
where $\Gamma_\perp = 2 \pi v^2_0 N(0) $.  For a finite
$\delta k$, $\rho_{s,imp}^c $ behaves similarly to the $\delta
k \rightarrow 0$ case, but the overall amplitude of
$\rho^c_{s,imp}$ decreases with increasing $\delta k$.

The above discussion indicates that both the lifetime effect
and the impurity assistant hopping contribute a $T^2$ term
to $\rho^c_s$ at low temperature.  The $T^5$ behavior of
$\rho_s^c$ from the coherent tunneling of electrons is
therefore observable only when $T_c\gg T \gg {\rm
max}(\Gamma , T^*)$, where $T^* \sim {1\over 2}
[\Gamma_\perp \Delta_0 / (t^0_\perp)^2]^{1/3}T_c$ is a
characteristic temperature above which the contribution from
the coherent hopping of electrons to $\rho_s^c$ is larger
than that from the impurity assisted hopping.  For a highly
anisotropic and clean high-$T_c$ superconductor, generally
$\Gamma_\perp\ll t_\perp^0$ and $t_\perp^0$ is smaller than
or of the same order as $\Delta_0$.  However, because of the
$1/3$ exponent in $T^*$, $T^*$ is a rather large temperature
scale compared with even $\Gamma$.  Thus assisted
hopping may affect the $c$-axis superfluid response in a
larger temperature region than the scattering lifetime
effect.

Summarising the discussion above, we may draw a qualitative phase
diagram for the temperature dependence of $\rho_s^c$ and
$\rho_s^{ab}$ at low temperature:  When $T\ll \Gamma$, the
lifetime effect is important, both $\rho^{ab}$ and
$\rho^c_s$ vary quadratically with $T$.  When $T^*\gg T \gg
\Gamma$, the impurity assisted hopping becomes more
important to the $c$-axis superfluid response.  In this
case, $\rho^c_s$ varies still quadratically with $T$, but
the coefficient of the $T^2$ term in $\rho^c_s(T) /
\rho_s^c(0)$ is proportional to $\Gamma_\perp
/(t_\perp^0)^2\Delta$, which is generally smaller than that
in the $T\ll \Gamma$ case where the coefficient of $T^2$
term in $\rho^c_s(T) / \rho_s^c(0)$ is proportional to
$\Gamma/ \Delta^3$.  When $T_c \gg T\gg \Gamma$ for
$\rho_s^{ab}$ or $T_c\gg T \gg T^*$ for $\rho_s^c$, the
disorder effect is not so important and the temperature
dependences of $\rho^c_s$ and $\rho^{ab}_s$ in a clean
system should be recovered.  Thus there is a crossover from
$T^2$ to $T$ when $T\sim \Gamma$ for $\rho^{ab}_s$ and from
$T^2$ to $T^5$ when $T\sim T^*$ for $\rho^c_s$ as $T$
increases at low temperature\cite{note}. 
A simple formula which gives a qualitative measure 
for the crossover of $\rho_s^c$ from the intrinsic 
$T^5$ behavior to the disorder $T^2$ behavior is 
$\rho_s^c (T)\approx \rho_s^c(0)+\alpha T^2 
[1+(T/T^\star )^3]$ with $\alpha$ a scattering potential 
dependent constant. The above picture may be also 
valid for the case of substitutional impurities such 
as Zn, but we cannot rule out a possible formation of a 
``impurity band'' which may have a stronger effect 
on the low temperature behavior of $\rho_s^c$ than 
the disorder effects previously discussed.

In comparing the predictions of this model to experiment for
a given system, it is essential to assess whether coherent
Bloch band are present at all.  The answer to this question
in the normal state of highly anisotropic copper-oxides,
such as BSSCO, appears to be no. The $c$-axis mean free
path is generally very short compared with the $c$-axis 
lattice constant.  Microwave\cite{Bonn92}
and thermal Hall measurements\cite{Krishana95}, however,
reveal that the mean free path grows very rapidly in the
superconducting state.  The superconducting quasiparticle
scattering rate is therefore much lower than the
extrapolated normal state scattering rate.  Thus it is our
belief that the available experimental data favor coherent
Bloch band formation along $c$-axis at low temperature below
$T_c$.  For BSSCO, $t_\perp^0$ is small and $\rho_s^c$ might be
dominated by the impurity assisted hopping\cite{Radtke96}.
However, for other compounds with lattice tetragonal
symmetry and relatively lower anisotropy, such as LaSrCuO,
TlBaCaCuO and HgBaCuO, we believe that the coherent band
should have substantial contribution to $\rho_s^c$.
Experimentally, a crucial test for this can be done by
carefully measuring and analysing $\rho_s^c$ data in these
compounds to see if the predicted $T^5$ behavior exists.
(The $T^5$ behavior in $\rho_s^c$ is not applicable to YBCO
as it has CuO chain layers and non-flat CuO
planes\cite{xiang96}.)

In conclusion, we have demonstrated that there exists a
remarkable interplay between underlying electronic structure
and orbital parameter symmetry in Copper-Oxides.  The
$c$-axis electromagnetic response reflects not only the
symmetry of the order parameter but also the relative
symmetry of the Copper orbitals involved in $c$-axis and
$ab$-plane conduction.  At low temperature the disorder
effect dominates and both $\rho_s^c$ and $\rho_s^{ab}$ have
$T^2$ dependence.  Above some characteristic temperatures,
which may be different for $\rho_s^c$ and $\rho_s^{ab}$, the
temperature dependences in a clean system, linear $T$ for
$\rho_s^{ab}$ and $T^5$ for $\rho_s^c$, are expected.

The authors are grateful for valuable discussions with 
D. C. Morgan, N.  E.  Hussey, and J. R. Cooper of  
the IRC in Superconductivity at Cambridge.



\begin{references}
\bibitem{srep} Another choice is bonding and non-bonding O
Wannier states with respect to the Cu $s$ orbital.  In this
representation the $d$ orbital (instead of the O bonding
orbital) has hybridisation matrix elements $\mu_k$ and $z_k$
with respect to both bonding and anti-bonding orbitals,
respectively.
\bibitem{ab-plane} The $ab$-plane hopping integral also
contains a term proportional to $\mu_k^2$ derived from
effective O-O hopping, which is often parameterised by
$t^\prime$ and $t^{\prime \prime}$ hopping.
\bibitem{note} It is also possible that $T^*$ is very large 
which washes out completely the $T^5$-law regime.
\end{references}
\end{document}